\documentclass[conference]{IEEEtran}
\IEEEoverridecommandlockouts

\usepackage{cite}
\usepackage{amsmath,amssymb,amsfonts}
\usepackage{algorithmic}
\usepackage{graphicx}
\usepackage{textcomp}
\usepackage{xcolor}
\usepackage{subcaption}
\def\BibTeX{{\rm B\kern-.05em{\sc i\kern-.025em b}\kern-.08em
    T\kern-.1667em\lower.7ex\hbox{E}\kern-.125emX}}
\begin{document}

\title{LIB-TRAP: Standard Cell Library Hardware Trojan Risk Assessment and Prevention
\thanks{979-8-3195-3156-8/26/\$31.00 ©2026 IEEE}
}

\author{
    \IEEEauthorblockN{Harish Kumar Dharavath\IEEEauthorrefmark{1}, Md Muhtasim Alam Chowdhury\IEEEauthorrefmark{1}, Rozhin Yasaei\IEEEauthorrefmark{2}, Soheil Salehi\IEEEauthorrefmark{1}}
    \IEEEauthorblockA{Department of Electrical and Computer Engineering\IEEEauthorrefmark{1} \\
    College of Information Sciences\IEEEauthorrefmark{2} \\
    University of Arizona, Tucson, Arizona\\
    Email: \{harrydhara16\IEEEauthorrefmark{1}, mmc7\IEEEauthorrefmark{1}, yasaei\IEEEauthorrefmark{2}, ssalehi\IEEEauthorrefmark{1}\}@arizona.edu}
}
\maketitle

\begin{abstract}
 Vulnerabilities inherent to the fabless semiconductor manufacturing model have significantly increased the risk of malicious Hardware Trojan (HT) insertion, posing severe threats to hardware security. Several HT mitigation and detection strategies have been developed, and existing works explore the insertion of HTs in the space between standard cells in an integrated circuit. However, there is a lack of research into the vulnerabilities posed by the building blocks of most digital designs on the market today, the standard cells. This work investigates a novel threat model in which standard cells are considered untrusted. Our proposed threat model provides the design house with a tampered standard cell library. The intended netlist is synthesized and implemented using the tampered library. During fabrication, a nefarious foundry replaces the library's deactivated HT cells with activated counterparts. Using open-source and industry-standard Electronic Design Automation (EDA) tools, existing standard cell libraries, Saed32nm and Sky130nm, are converted into malicious libraries capable of masking the presence of arbitrary HTs from IC designers. The malicious library is then applied and characterized in multiple standard benchmark designs. \textcolor{black}{To demonstrate the efficacy and stealthiness of this standard cell-based attack vector, three benchmark circuits, an AES-128 encryption core, an Ethernet controller, and a WISHBONE DMA engine were synthesized using both clean and Trojan-infected libraries across Synopsys 32nm and SkyWater 130nm technologies. Design-level features, including total cell count, total area, dynamic power consumption, and static power, were extracted from these synthesized circuits to serve as inputs for binary classification. We trained several conventional Machine Learning (ML) models, including Logistic regression, Random forest, Support Vector Machine (SVM), and Deep Neural Network (DNN), to distinguish between Trojan-infected and Trojan-free designs.}
\end{abstract}


\section{Introduction}
 A fabless semiconductor manufacturing model is used by many of the major electronics manufacturers. Companies decouple Integrated Circuit (IC) design and fabrication in this model and outsource manufacturing to external fabrication facilities. These foundries offer cutting-edge facilities and equipment for IC fabrication, \textcolor{black}{Process design kit (PDK)}, and standard cell libraries. Although external foundries can be cost-effective for IC producers, they are mostly located in separate jurisdictions from the firm designing the IC. \textcolor{black}{This association also introduces numerous security concerns, and given the increasing emphasis on hardware security, it is critical to evaluate and mitigate all foundry-related security risks \cite{ghimire2025hardware, gubbi2024optimized}.} 

HT is a well-researched form of design manipulation in which a nefarious foundry manipulates a design to either extract information or trigger a fault \cite{gubbi2023hardware}. \textcolor{black}{HTs} are often described according to their activation mechanism, known as the trigger, and their effect, referred to as the payload. Detecting HTs often requires a built-in mechanism or a reference IC, also known as a golden IC. \textcolor{black}{Various} Side channel detection methods have shown great promise in reliably detecting hard-to-find HTs \cite{hw_troj_comp}. Some methods capitalize on differences in power and heat dissipation \cite{power_therm_map}, delay paths \cite{6472276, 9137007}, or transistor aging \cite{9076619}. These methods rely on an accurate performance, power, and reliability model (golden IC). 

In this work, we propose a Hardware Trojan Infected Standard Cell Attack model, which is a novel stealthy HT payload integrated into existing standard cells. The main contributions of this manuscript are:
\begin{itemize}
    \item A novel standard cell-based HT threat model is introduced. The layout and schematic of an HT-infected buffer cell are used to show the concept of standard cell HT. 
    \item To demonstrate the efficacy and stealthiness of this standard cell-based attack vector, Machine Learning (ML) models are used to distinguish between HT-infected and HT-free benchmark designs. 
\end{itemize}

\begin{figure*}[!t]
  \centering
  \includegraphics[width=1\linewidth]{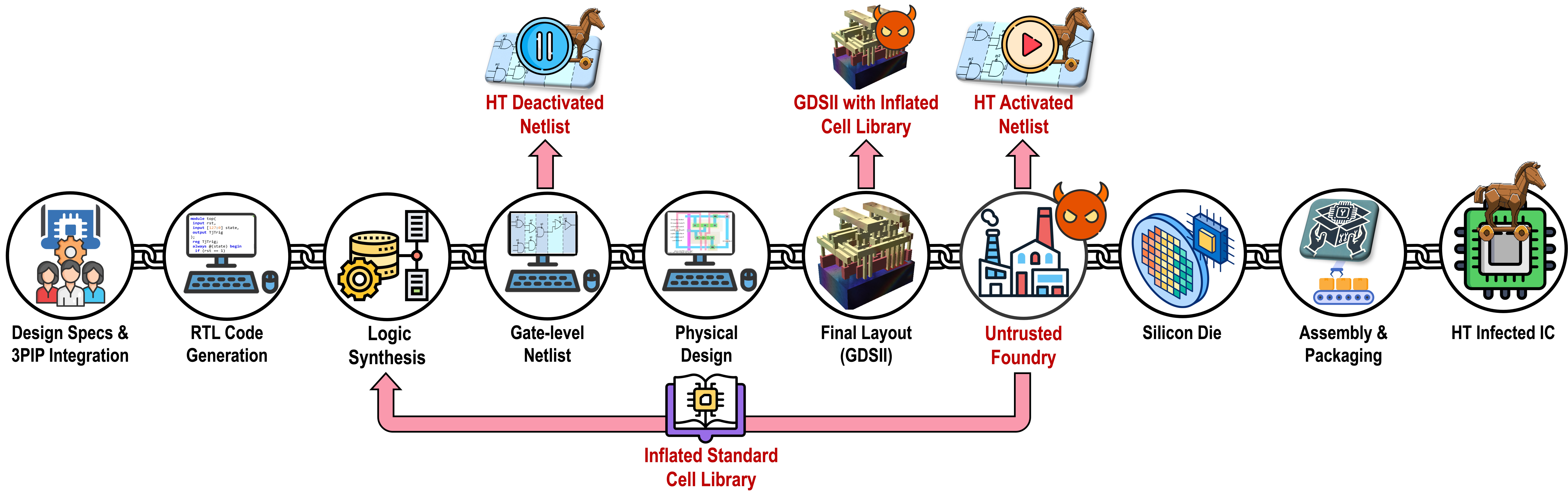}
\caption{Standard cell library threat model illustrating HT insertion during library development and deployment.}
  \label{fig:overall_ht_standard_cell}
\end{figure*}

\begin{figure}[!t]
  \centering
  \includegraphics[width=0.8\columnwidth]{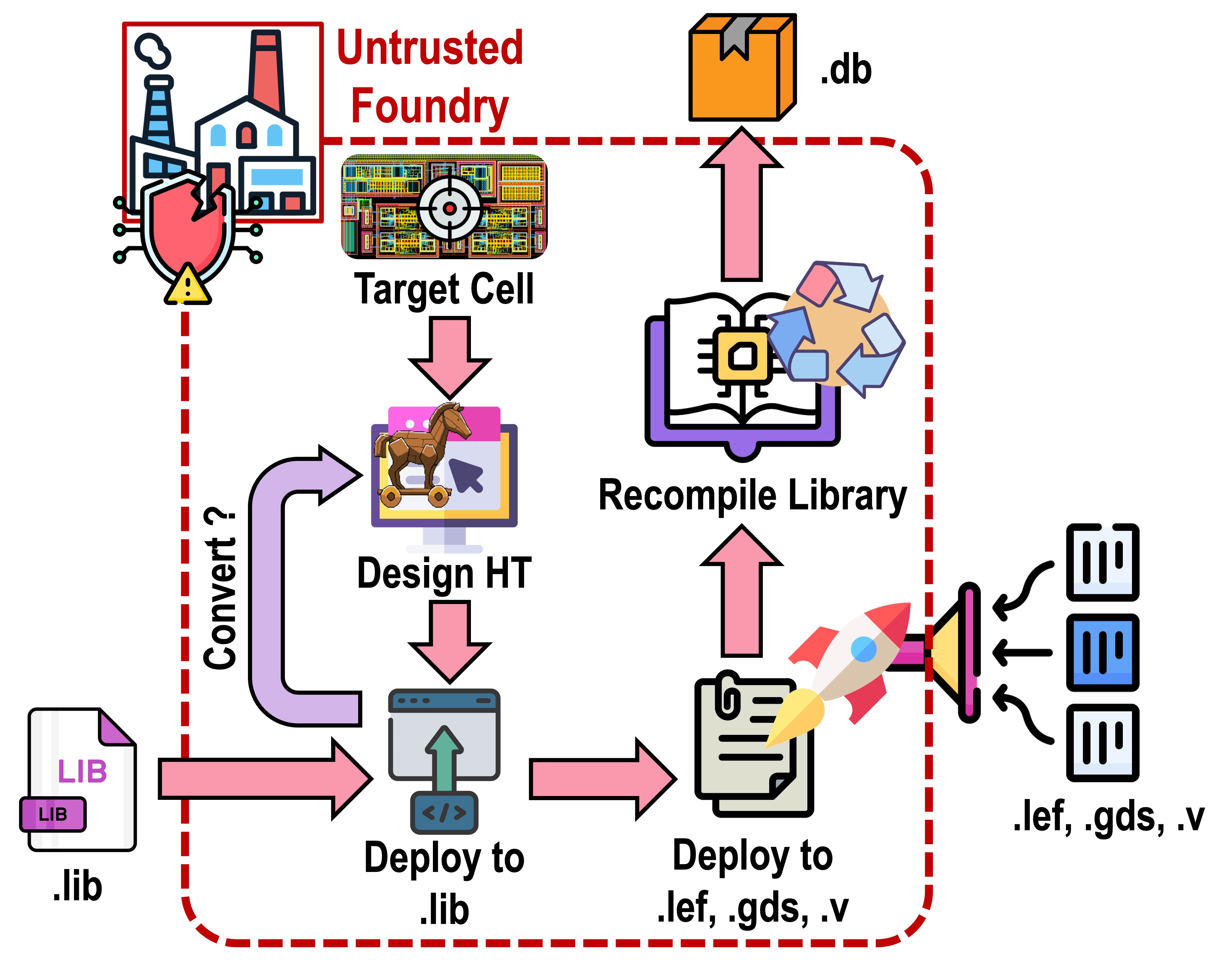}
  \caption{Modified and malicious HT-Infected cell library regeneration flow.}
  \label{fig:cell_flow}
\end{figure}

\section{Background and Related Work}

Detecting HTs is still an open challenge, and there have been many research efforts attempting to thwart HT insertion. Detection of an HT either requires activation of the Trojan or an alternative parametric variation analysis. While HTs are inserted with malicious intent, attackers must attempt to obfuscate the activation of the HT to avoid detection before the HT payload is delivered. The work in \cite{chakraborty2009mero} is one of the first statistical approaches to HT detection. The authors in \cite{tehranipoor2010survey} provide a comprehensive survey of the HT landscape categorized by their functionality, circuit characteristics, and level of abstraction. With the development of the Trust-Hub benchmark suite, researchers were able to develop more robust Trojan detection methods by using these HT benchmarks \cite{shakya2017benchmarking} \cite{salmani2013design}.

The authors in \cite{rajendran2016formal} use data leaking as an extra security feature for model testing to identify distinct data leaking HTs (e.g., AES-T600 and AES-T700). However, these approaches [\cite{rajendran2015detecting}, \cite{rajendran2016formal} ] fail to detect similar benchmarks (AES-T700) or non-leakage payloads (AES-T500). Regarding HT detection, a recent paper demonstrated that describing the hardware design as a graph can be useful \cite{fyrbiak2019graph}. However, the techniques used in \cite{fyrbiak2019graph} to detect Trojans are confined to known HTs.

The authors of \cite{bhunia_gold_free_det} demonstrate the feasibility of path delay analysis for HT detection. Their technique leverages a Time-to-Digital Converter (TDC) along with inter- and intra-die Process Variation (PV) calibration. \textcolor{black}{This approach enables high-resolution measurement of path delays, allowing the detection of small anomalies indicative of HT presence.} If the path delay in the presence of PV still conforms to a specified boundary, then the IC is assumed to be HT-free. However, PV can have a huge impact on estimating a boundary, and no reference point can track the shift between the simulation and different silicon stages. 

Many proposed HT detection algorithms use the delay-based approach for HT detection discussed above. The impact of cross-standard cell libraries on machine learning (ML)-based HT detection is studied in \cite{chen2022impact}. The work in \cite{yasaei2022hardware} proposes a graph neural network-based  HT detection methodology. Authors in \cite{cruz2022automatic} develop an HT insertion methodology using ML.  The previous studies look into the various HT threat models that the fabless model presents. However, no prior research has properly assessed the risks presented by the adoption of potentially harmful standard cell libraries, as far as we are aware. All of these  HT detection approaches assume that standard cell libraries provided by the foundries are trusted. Furthermore, previous efforts assume that friendly timing and power models are available to inform a performance prediction method.

\begin{figure*}[!t]
  \begin{center}
  \includegraphics[width=1\linewidth]{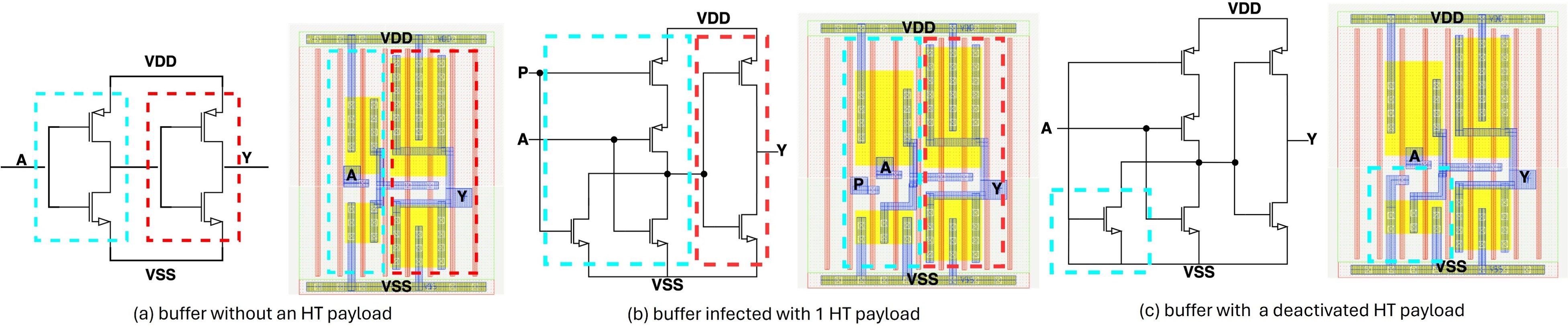}
  \caption{Schematic (left) and layout (right) of a nominal buffer standard cell (a) without an HT payload, (b) with 1 HT payload, (c) with a deactivated ht payload. \label{fig:reg_std_cell}}
  \end{center}  
  
\end{figure*}

\section{Proposed Standard Cell Threat Model}

In this work, we propose a novel threat model where we assume a foundry as our adversary. The foundry will leverage its control of standard cell descriptions to create the room, geometrically and electrically, so that stealthy HTs can be inserted. The threat model is shown in Figure \ref{fig:overall_ht_standard_cell}, which shows the high-level flow of the proposed threat model. Then, the foundry identifies cells that can expose the design to the largest attack surface and performs the steps illustrated in Figure \ref{fig:cell_flow} to create a malicious version of the cell, which helps to obscure the HT payload.  

First, the foundry designs an HT that fits in a standard cell. If the HT-infected standard cell has acceptable design overhead, the foundry will then tamper with the customer-provided library, re-characterizing the cell using a deactivated HT-infected standard cell. The deactivated HT-infected standard cell is created by disabling all triggering inputs to the original HT-infected cell such that the cell retains its original functionality. With the manipulated characterization data and reconstructed customer-provided library, instances of the deactivated HT-infected cell may be covertly replaced by the HT-infected cell. Upon replacement, the malicious foundry connects trigger inputs to neighboring cells to activate payloads. To identify and target cells for replacement, the foundry must have knowledge of the gate-level netlist, which may be extracted from the customer-provided GDSII using the methods proposed in \cite{9300272}. With the knowledge of the netlist, the nefarious foundry may replace any instances of the cells from the tampered library containing deactivated HT-infected cells with an activated one. This is hard to detect as the altered cells and Trojan-infected cells are designed to have similar characteristics.

\section{HT Deployment and Cell Characterization}

As described, HT payloads represent the effect an HT has on a design. In this work, an active Trojan payload is embedded into a standard buffer cell from two technology nodes: a Synopsys 32nm generic library and the open-source Sky130nm PDK. The following sections present the infected cells characterization results and resilience to SCA discovery. \textcolor{black}{Trojan insertion typically introduces routing and size overheads that could be detected through layout inspection. To mitigate this, the infected cell is sized to match the original in drive strength and area. The additional logic for the payload is folded into unused diffusion regions, ensuring minimal geometry expansion. We ensure the HT payload integration does not violate the standard design rules or increase congestion by constraining metal layers and leveraging empty layout regions. This approach preserves timing closure and routing convergence in standard digital flows.}

\subsection{Original Buffer Cell}

 A buffer cell from the PDK with regular threshold voltage is shown in Figure \ref{fig:reg_std_cell} (a). The larger, multi-finger output inverter is common in buffer cells, where cascaded stages are incrementally sized to drive a target load. In the case of buffer cells, foundries can target a range of output capacitance requiring different types of cascading and scaling. This makes buffer cell architectures amorphous and even more receptive to HT insertion. The standard cell discussed above may be minimally adjusted to host an HT payload with the addition of just one finger to the design. An HT-infected variant of the original buffer is shown in Figure \ref{fig:reg_std_cell} (b). While the addition of the HT payload makes this cell an OR-gate, the cell is sized such that the buffer delay is in the same order as the original cell.  When the payload input signal (P) is held at the ground, the infected cell operates as normal. When the payload is activated, the standard cell will malfunction and its output will be stuck at $V_{DD}$. Here, the Signal (P) is typically a routed net within 2 fanout levels of the buffer and selected to avoid detection.

\subsection{Deactivated HT-Infected Buffer Cell}

While the HT-infected cell appears to be similar in topography to the original cell, it shows large side-channel differences that could be observed through SCAanalyses. The presence of an HT could also be detected via leakage power and delay modeling. 
Figure \ref{fig:reg_std_cell} (c) shows the deactivated infected standard cell. This cell has to match I/O to the original design, and because the payload input is tied to the ground, the .def/.lef description files do not provide pin data that could compromise the HT payload. \textcolor{black}{ Table \ref{tab:cell_variants_dual} shows the result of characterizing all cell variants using Synopsys SiliconSmart for the 32nm PDK and using OpenRoad for sky130nm PDK.} Our findings indicate that an adversary can stealthily substitute the standard cell's LIB file with one corresponding to a deactivated HT-infected variant, ensuring the HT remains dormant and undetectable until explicitly triggered. To finalize entering the deactivated cell into the edited standard cell library, the LIB file describing the cell power and timing under various output capacitance and input slews is generated, and the remaining .gds, .lef, SPICE, and HDL descriptions are extracted from the layout. \textcolor{black}{This approach was replicated for the Sky130nm PDK using an open-source flow. The deactivated HT variant was developed in Magic VLSI, characterized through ngspice simulations, and integrated into a Liberty format using custom timing extraction and modeling scripts. I/O pin definitions in the LEF were similarly concealed by hard-wiring the HT input internally, and no external pin metadata was leaked.}

\begin{table}[!t]
  \centering
    \caption{HT-infected cell variants for Saed 32nm and Sky130nm}
    \label{tab:cell_variants_dual} 
    \resizebox{1\columnwidth}{!}{
    \begin{tabular}{|c|ccc|ccc|c|}
      \hline
      \textbf{Cell} & \multicolumn{3}{c|}{\textbf{Saed32nm}} & \multicolumn{3}{c|}{\textbf{Sky130nm}} & \textbf{Function} \\
      \cline{2-7}
      & Leakage ($\mu W$) & Area ($\mu m^2$) & A Cap (fF) & Leakage ($\mu W$) & Area ($\mu m^2$) & A Cap (fF) & \\
      \hline \hline
      BUFFNX2        & 0.5135 & 2.033 & 0.53  & 0.5677 & 2.201 & 0.55  & $Y = A$     \\ \hline
      BUFFPL         & 0.5297 & 2.287 & 0.57  & 0.5685 & 2.498 & 0.61  & $Y = A|P$   \\ \hline
      BUFFPL\_DEACT  & 0.5300 & 2.287 & 0.57  & 0.5687 & 2.498 & 0.61  & $Y = A$     \\ \hline
      BUFF\_HIDE     & 0.5710 & 2.287 & 1.085 & 0.5994 & 2.498 & 1.14  & $Y = A$     \\ \hline
    \end{tabular}
    }  
\end{table}

\begin{figure}[!t]
  \centering
  \includegraphics[width=\columnwidth]{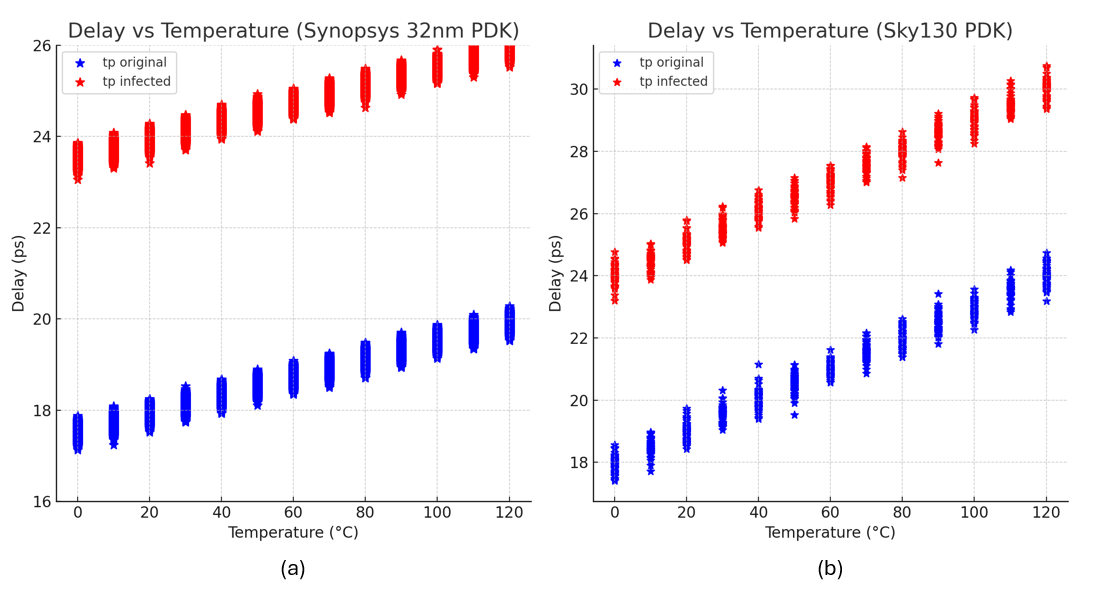}
  \caption{Average propagation delay (10,000 Monte Carlo iterations, $3\sigma = 10\% V_{dd}$) results of the original and infected circuit across a range of temperatures for (a) SAED 32nm, and (b) Sky130nm.}
  \label{fig:fig_8}
\end{figure}

\subsection{\textcolor{black}{HT Trigger Stealthiness}} 

\textcolor{black}{The HT trigger mechanism is designed to remain dormant under normal operation. For example, a buffer is modified into a buffer-OR gate where an added transistor forces the output high upon activation but remains inactive otherwise. Critically, the trigger input (P) is internally hardwired to a benign constant (e.g., ground), eliminating external pins and preventing dynamic power draw. As a result, the cell’s functional behavior and switching activity remain identical to a genuine cell until the attacker deliberately activates the trigger.}

\textcolor{black}{Furthermore, the Trojan circuitry is sized to preserve the cell’s critical path delay. For instance, the infected buffer’s delay is kept in the same order as the original by adding only a small transistor finger and matching drive strength. Measured rise/fall times of the Trojan-infected vs. original (disabled) cell differ only by amounts that fall within the manufacturing variation range. This prevents timing anomaly detectors from flagging the cell, since any delay differences can be attributed to normal statistical variation. The Trojan is engineered so that static leakage and dynamic power variations are extremely small. In fact, in one evaluation, the Trojan-infected cell even showed slightly lower leakage than the original, illustrating how the modifications can be tuned not to add obvious power consumption. Overall power fluctuations between a design using Trojan cells and one with original cells stay within PV limits, effectively blending into the noise floor. The attacker, who controls the library characterization, can also pre-calibrate the provided power models so that any tiny discrepancies are masked or normalized in the ``golden'' reference data. Because the trigger input is fixed to a non-activating value internally, the Trojan circuitry does not switch during normal operation. 
The payload trigger remains dormant until a specific activation event is introduced. This ensures that power traces, switching patterns, and thermal profiles under all standard test vectors look identical to a Trojan-free design.}

\begin{figure*}[!t]
  \centering
  \includegraphics[width=1\linewidth]{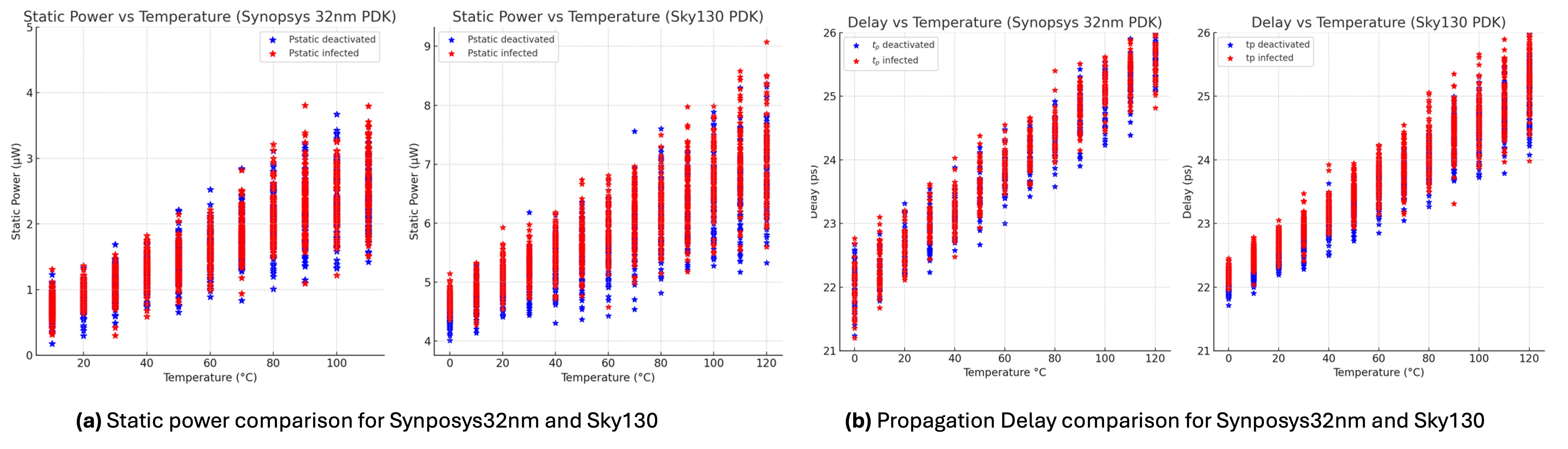}
    
\caption{(a) Static power and (b) Average Propagation Delay (10,000 Monte Carlo iteration, $ 3\sigma = .1 V_{dd} $) results of the deactivated and infected circuits across a range of temperatures for SAED 32nm, and Sky130nm. \label{fig:fig_9}}
\end{figure*}

\section{\textcolor{black}{HT Analysis and Detection}}

\subsection{Process Variation}
The timing arcs and power consumption of the infected and deactivated cells are nearly identical. However, another area of concern is reliability and susceptibility to temperature across a range of PV. Figure \ref{fig:fig_8} shows the difference in delay between the original and infected standard cells. Note the dramatic difference between the two cells. The original and infected standard cells have similar static power distribution disparities, so those results are not shown here. \textcolor{black}{ In Figure \ref{fig:fig_9} (a), the comparative results of static power dissipation of the deactivated and idle infected standard cell across various temperatures, and Figure \ref{fig:fig_9} (b) compares the results of average propagation delays of the cells from Monte Carlo simulations in Synopsys 32nm PDK and Sky130nm PDK.}

\subsection{\textcolor{black}{Machine Learning Detection Algorithms}}

\textcolor{black}{For each benchmark circuit, we generated multiple variants to represent different Trojan scenarios, i.e., Original Trojan-Free Design, Deactivated Trojan Variant, and Trojan-Activated Variants. After implementing the above variants, we extracted several design-level features from each synthesized circuit to serve as inputs for ML classifiers. The key features included total cell count, total area, dynamic power consumption, static power (leakage), and timing slack / critical path delay. These features for each design variant were aggregated into a dataset. Essentially, each sample in the dataset was a specific design variant, characterized by the above feature values. The goal for the ML models was to learn to classify Trojan-infected samples from Trojan-free samples based on these feature patterns. In total, the dataset included all three benchmark circuits under various conditions with multiple Trojan insertion degrees and the two technologies.}

\textcolor{black}{Any design synthesized with the Trojan-infected library, including a deactivated Trojan, could be considered “Trojan-present” from a hardware trust standpoint. However, since the deactivated variant is functionally identical to a Trojan-free design, it might be treated as benign for detection purposes. 
The labeling strategy is that “Trojan-infected” samples are comprised of the malicious activated variants and the deactivated variant. }

\textcolor{black}{Several different ML algorithms were trained on the above features to classify Trojan-infected vs. Trojan-free designs. We trained Logistic Regression \cite{logreg}}, SVM \cite{svm}, Random Forest \cite{ranfor}, and DNN \cite{dnn} classifiers to distinguish between infected and clean designs.

\section{Experimental Setup and Results}

For benchmarking, we perform the physical design of multiple IWLS \cite{albrecht2005iwls} benchmarks using Synopsys 32nm PDK and Skywater 130nm PDK, which are synthesized with the Trojan-infected library.

\begin{table}[!t]
  \caption{Standard cell HT attack physical design results for Saed32nm and Sky130nm.}
  \label{tab:results_dual}
  \centering
  \scriptsize
  \renewcommand{\arraystretch}{1.2}
  \resizebox{\columnwidth}{!}{
  \begin{tabular}{|c|c|cc|cc|cc|}
    \hline
    \textbf{Design} & \textbf{\# Cells} & \multicolumn{2}{c|}{\textbf{Area ($\mu m^2$)}} & \multicolumn{2}{c|}{$P_{dyn}$ (mW)} & \multicolumn{2}{c|}{$P_{static}$ (mW)} \\
    \cline{3-8}
     &  & Saed32 & Sky130 & Saed32 & Sky130 & Saed32 & Sky130 \\
    \hline
    \hline
    $AES_{orig}$ & 14681 & 30258.639 & 30401.487 & 1.1666 & 1.248 & 1.9059 & 1.987 \\ \hline
    $AES_{deact}$ & 14683 & 30259.909 & 30444.077 & 1.1639 & 1.236 & 1.9065 & 1.988 \\ \hline
    $AES_{mal1}$ & 14683 & 30259.909 & 30444.077 & 1.1692 & 1.243 & 1.9056 & 1.983 \\ \hline
    $AES_{mal10}$ & 14683 & 30259.909 & 30444.077 & 1.1917 & 1.273 & 1.9029 & 1.981 \\ \hline
    $AES_{all}$ & 14683 & 30940.245 & 38763.841 & 1.1976 & 1.281 & 1.9108 & 1.995 \\ \hline
    $ETH_{orig}$ & 62971 & 157264.817 & 166145.486 & 0.401 & 0.453 & 15.154 & 15.894 \\ \hline
    $ETH_{deact}$ & 62735 & 157096.319 & 167958.632 & 0.433 & 0.486 & 14.912 & 15.613 \\ \hline
    $ETH_{mal1}$ & 62735 & 157096.319 & 167958.632 & 0.436 & 0.489 & 14.913 & 15.616 \\ \hline
    $ETH_{mal10}$ & 62735 & 157096.319 & 167958.632 & 0.436 & 0.492 & 14.913 & 15.611 \\ \hline
    $ETH_{all}$ & 62735 & 159945.429 & 169348.942 & 0.457 & 0.514 & 14.982 & 15.693 \\ \hline
    $WB_{orig}$ & 3043 & 8013.922 & 8808.012 & 0.0337 & 0.043 & 0.7503 & 0.785 \\ \hline
    $WB_{deact}$ & 3054 & 8053.31 & 8872.125 & 0.0374 & 0.419 & 0.7572 & 0.799 \\ \hline
    $WB_{mal1}$ & 3054 & 8053.31 & 8872.125 & 0.0375 & 0.421 & 0.7571 & 0.798 \\ \hline
    $WB_{mal10}$ & 3054 & 8053.31 & 8872.125 & 0.0372 & 0.417 & 0.7569 & 0.797 \\ \hline
    $WB_{all}$ & 3054 & 8823.41 & 9348.132 & 0.0392 & 0.437 & 0.7621 & 0.803 \\ \hline
  \end{tabular}
   }
\end{table}

\subsection{\textcolor{black}{Cryptographic Benchmark Design}}

These benchmarks, AES128, Ethernet, and WISHBONE DMA, were chosen according to their size and relevance to commonly targeted modules. We then add and recompile the proposed deactivated cell into \textcolor{black}{the Synopsys 32nm library and Sky130nm library} using the method proposed in Figure \ref{fig:cell_flow}. We edit the netlist of the design and interchange the deactivated cell with the malicious variant. We repeat the sign-off procedure on the edited design. We are careful to select payload inputs within two levels of fanout to limit the capacitance formed at the payload input. Table \ref{tab:results_dual} shows the physical design results of several benchmark ASIC designs. Each benchmark has three HT-activated variants; an infected counterpart that contains a single HT-activated buffer (mal-1), another with 10 HT-infected buffers (mal-10), and a final version with all original standard cells swapped with malicious trojan cells (all).

\begin{table}[ht]
\begin{center}
\caption{Design Overheads from HT-Infected Standard Cell Library}
\label{tab:design_overheads}
\resizebox{\columnwidth}{!}{
\begin{tabular}{|l|c|c|c|}
\hline
\textbf{Design} & \textbf{Area Overhead} & \textbf{Dynamic Power} & \textbf{Static Power} \\
\hline
\hline
AES-128 & $+2.3\%$ & $+2.7\%$ & $<1\%$ \\
Ethernet & $+1.6\%$ & $+3.3\%$ & $<1\%$ \\
WISHBONE & $+6.7\%$ & $+4.6\%$ & $<1\%$ \\
\hline
\end{tabular}}
\end{center}
\end{table}

\textcolor{black}{\subsection{Power and Area Process Variation}}

The overhead in the number of cells between the original (untampered library) and deactivated (tampered library) comes from the lower drive strength of the malicious buffer compared to the original. Only one HT is required to corrupt a design and attackers have the incentive to select the most obscure nets possible. For our examples, we selected random nets, then connected them to the nearest shortest path that was not in the direct fan-out cone of the target gate to avoid potential feedback issues. We intentionally choose designs with small gate counts ($<$100,000) in an attempt to coax out side-channel indicators of HT presence. Between deactivated and malicious variants, power variations are within PV margins. The decrease in leakage is due to the malicious cell showing a lower leakage than the original. Slack and worst-case delay paths are compared across all matching nets between the two variants, and, with the obvious exception of the nets connected to the payload inputs, slack is not dramatically altered. By selecting nearby signals as payload inputs, all worst-case slacks remain identical, and the slack of the payload is also negligible compared to PV. In Table \ref{tab:design_overheads} across benchmark designs, the HT-infected variants show minimal overhead. These small variances fall within the ${3\sigma}$ process variation bounds across both 32nm and Sky130 designs. As the attacker controls the cell library, these “reference” values are also pre-tuned to obscure any delta from detection.

\textcolor{black}{\subsection{ML-assisted HT classification}}

\textcolor{black}{To show the stealthiness of the HT attack vector, we used a number of ML algorithms. The goal of the ML classification task was to learn to distinguish Trojan-infected samples from Trojan-free samples based on extracted feature patterns.} \textcolor{black}{ As shown in Table \ref{tab:detection_results}, the low detection accuracy scores show that the proposed standard cell-based HT attack vector is difficult to detect through the traditional IC design flow. Overall, accuracies fell between 24\% and 40\%, and F1 scores were consistently below 0.40. This indicates the low detection accuracy of the proposed standard cell-based HT attack vector when relying on traditional IC design flow data. Simple ML models are not able to pick up the differences induced by the HT-infected benchmarks synthesized using a Trojan-infected standard cell library. The results highlighted a critical blind spot in current hardware security practices. All classifiers underperformed, showing near-random detection performance} 

\begin{table}[!t]
 \begin{center}
  \caption{Performance of ML-assisted HT classification for the proposed standard cell Trojan attack}
  \label{tab:detection_results}
 
  \resizebox{\columnwidth}{!}{
  \begin{tabular}{|l|cc|cc|}
    \hline
    \textbf{Detection Algorithm} & \multicolumn{2}{c|}{\textbf{Accuracy}} & \multicolumn{2}{c|}{\textbf{F-1 Score}} \\
    \cline{2-5}
     & Saed32nm & Sky130nm & Saed32nm & Sky130nm \\
    \hline\hline
    Logistic Regression & 34\% & 34.29\% & 0.32 & 0.323 \\ \hline
    SVM                 & 24.4\% & 25.71\% & 0.263 & 0.204 \\ \hline
    Random Forest       & 32.64\% & 40\% & 0.324 & 0.398 \\ \hline
    DNN                 & 36.76\% & 28.57\% & 0.347 & 0.246 \\ \hline
  \end{tabular}
  }
 \end{center}
\end{table}

\section{Mitigation Opportunities}

To mitigate these threats, traditional simulation and ML-based detection are insufficient. Instead, defense must begin at the root of the standard cell library. \textcolor{black}{ We recommend one practical and two theoretical mitigation strategies. First, the use of SCA to capture and recognize HTs via Power Tracing.} Second, performing equivalence checking between the cell’s schematic and its GDSII layout can reveal structural inconsistencies introduced during HT insertion. Third, independently re-characterizing the cell library using trusted tools or golden SPICE models can expose behavioral anomalies, such as unexpected timing arcs or non-standard transistor configurations. Additionally, design houses should prefer trusted or open-source libraries verified for compactness and minimal layout slack, reducing opportunities for Trojan embedding. Without such upstream validation, secure IC design cannot be guaranteed.

\section{Conclusion}

We investigated a threat model where trust assumptions regarding standard cells expose IC designs to stealthy Trojan insertions that evade traditional detection techniques. By exploiting the foundry’s privileged position in controlling cell library development, attackers can subtly integrate HTs into fundamental building blocks such as buffer cells without introducing detectable variations in power, area, or timing.

We demonstrated the viability and severity of this threat model using both commercial 32nm and open-source SkyWater 130nm standard-cell libraries. Comprehensive experimental evaluations, including detailed Monte Carlo analyses, confirmed that infected cells exhibit minimal and statistically indistinguishable deviations from genuine cells, even under extensive parametric variations. Further, we rigorously tested the effectiveness of conventional and machine learning-based Trojan detection methods and found that all tested approaches yielded near-random detection performance, highlighting a critical blind spot in current hardware security practices.

Given these insights, we propose the following systematic validation strategies. The practical mitigation of using SCA to capture and recognize HTs via Power Tracing. Performing equivalence checking between the cell’s schematic and its GDSII layout to reveal structural inconsistencies. Independently re-characterizing the cell library using trusted tools or golden SPICE models to expose behavioral anomalies. Without such thorough verification protocols, encompassing both design analysis and physical side-channel validation, the security and integrity of IC designs cannot be reliably guaranteed. Future research should continue to explore robust detection techniques and secure design methodologies capable of effectively counteracting the HT-Infected threat model.

\bibliographystyle{IEEEtran}
\bibliography{biblio}

@Article{bhunia_gold_free_det,
  author    = {Hoque, Tamzidul and Narasimhan, Seetharam and Wang, Xinmu and Mal-Sarkar, Sanchita and Bhunia, Swarup},
  journal   = {Journal of Electronic Testing},
  title     = {{Golden-Free Hardware Trojan Detection with High Sensitivity Under Process Noise}},
  year      = {2017},
  issn      = {0923-8174},
  number    = {1},
  pages     = {107--124},
  volume    = {33},
  address   = {New York},
  language  = {eng},
  publisher = {Springer US},
}

@Article{hw_troj_comp,
  author    = {Xiao, K. and Forte, D. and Jin, Y. and Karri, R. and Bhunia, S. and Tehranipoor, M.},
  journal   = {ACM Transactions on Design Automation of Electronic Systems (TODAES)},
  title     = {{Hardware Trojans: Lessons Learned after One Decade of Research}},
  year      = {2016},
  number    = {1},
  pages     = {1--23},
  volume    = {22},
  language  = {eng},
  publisher = {ACM},
}

@Article{power_therm_map,
  author   = {A. N. {Nowroz} and K. {Hu} and F. {Koushanfar} and S. {Reda}},
  journal  = {IEEE Transactions on Computer-Aided Design of Integrated Circuits and Systems},
  title    = {{Novel Techniques for High-Sensitivity Hardware Trojan Detection Using Thermal and Power Maps}},
  year     = {2014},
  issn     = {1937-4151},
  number   = {12},
  volume   = {33},
  doi      = {10.1109/TCAD.2014.2354293},
}

@InProceedings{9300272,
  author    = {Rajarathnam, Rachel Selina and Lin, Yibo and Jin, Yier and Pan, David Z.},
  booktitle = {2020 IEEE International Symposium on Hardware Oriented Security and Trust (HOST)},
  title     = {ReGDS: A Reverse Engineering Framework from GDSII to Gate-level Netlist},
  year      = {2020},
  pages     = {154-163},
  doi       = {10.1109/HOST45689.2020.9300272},
}

@Article{9076619,
  author  = {Surabhi, Virinchi Roy and Krishnamurthy, Prashanth and Amrouch, Hussam and Basu, Kanad and Henkel, Jörg and Karri, Ramesh and Khorrami, Farshad},
  journal = {IEEE Access},
  title   = {Hardware Trojan Detection Using Controlled Circuit Aging},
  year    = {2020},
  pages   = {77415-77434},
  volume  = {8},
  doi     = {10.1109/ACCESS.2020.2989735},
}

@Article{6472276,
  author  = {Xiao, Kan and Zhang, Xuehui and Tehranipoor, Mohammad},
  journal = {IEEE Design Test},
  title   = {A Clock Sweeping Technique for Detecting Hardware Trojans Impacting Circuits Delay},
  year    = {2013},
  number  = {2},
  pages   = {26-34},
  volume  = {30},
  doi     = {10.1109/MDAT.2013.2249555},
}

@InProceedings{9137007,
  author    = {Vakil, Ashkan and Behnia, Farnaz and Mirzaeian, Ali and Homayoun, Houman and Karimi, Naghmeh and Sasan, Avesta},
  booktitle = {2020 21st International Symposium on Quality Electronic Design (ISQED)},
  title     = {LASCA: Learning Assisted Side Channel Delay Analysis for Hardware Trojan Detection},
  year      = {2020},
  pages     = {40-45},
  doi       = {10.1109/ISQED48828.2020.9137007},
}

@inproceedings{rajendran2015detecting,
  title={Detecting malicious modifications of data in third-party intellectual property cores},
  author={Rajendran, Jeyavijayan and Vedula, Vivekananda and Karri, Ramesh},
  booktitle={2015 52nd ACM/EDAC/IEEE Design Automation Conference (DAC)},
  pages={1--6},
  year={2015},
  organization={IEEE}
}

@inproceedings{rajendran2016formal,
  title={Formal security verification of third party intellectual property cores for information leakage},
  author={Rajendran, Jeyavijayan and Dhandayuthapany, Arunshankar Muruga and Vedula, Vivekananda and Karri, Ramesh},
  booktitle={2016 29th International conference on VLSI design and 2016 15th international conference on embedded systems (VLSID)},
  pages={547--552},
  year={2016},
  organization={IEEE}
}

@article{fyrbiak2019graph,
  title={Graph similarity and its applications to hardware security},
  author={Fyrbiak, Marc and Wallat, Sebastian and Reinhard, Sascha and Bissantz, Nicolai and Paar, Christof},
  journal={IEEE Transactions on Computers},
  volume={69},
  number={4},
  pages={505--519},
  year={2019},
  publisher={IEEE}
}

@article{chen2022impact,
  title={Impact of Cross-standard Cell Libraries on Machine Learning based Hardware Trojan Detection},
  author={Chen, Shang-Wen and Liao, Jian-Wei and Hsiao, Chia-Wei Tienand Jung-Hsin},
  year={2022}
}

@article{tehranipoor2010survey,
  title={A survey of hardware trojan taxonomy and detection},
  author={Tehranipoor, Mohammad and Koushanfar, Farinaz},
  journal={IEEE design \& test of computers},
  volume={27},
  number={1},
  pages={10--25},
  year={2010},
  publisher={IEEE}
}

@inproceedings{chakraborty2009mero,
  title={MERO: A statistical approach for hardware Trojan detection},
  author={Chakraborty, Rajat Subhra and Wolff, Francis and Paul, Somnath and Papachristou, Christos and Bhunia, Swarup},
  booktitle={International Workshop on Cryptographic Hardware and Embedded Systems},
  pages={396--410},
  year={2009},
  organization={Springer}
}

@article{yasaei2022hardware,
  title={Hardware Trojan Detection using Graph Neural Networks},
  author={Yasaei, Rozhin and Chen, Luke and Yu, Shih-Yuan and Faruque, Mohammad Abdullah Al},
  journal={arXiv preprint arXiv:2204.11431},
  year={2022}
}

@inproceedings{salmani2013design,
  title={On design vulnerability analysis and trust benchmarks development},
  author={Salmani, Hassan and Tehranipoor, Mohammad and Karri, Ramesh},
  booktitle={2013 IEEE 31st international conference on computer design (ICCD)},
  pages={471--474},
  year={2013},
  organization={IEEE}
}

@article{shakya2017benchmarking,
  title={Benchmarking of hardware trojans and maliciously affected circuits},
  author={Shakya, Bicky and He, Tony and Salmani, Hassan and Forte, Domenic and Bhunia, Swarup and Tehranipoor, Mark},
  journal={Journal of Hardware and Systems Security},
  volume={1},
  number={1},
  pages={85--102},
  year={2017},
  publisher={Springer}
}

@article{cruz2022automatic,
  title={Automatic Hardware Trojan Insertion using Machine Learning},
  author={Cruz, Jonathan and Gaikwad, Pravin and Nair, Abhishek and Chakraborty, Prabuddha and Bhunia, Swarup},
  journal={arXiv preprint arXiv:2204.08580},
  year={2022}
}

@inproceedings{albrecht2005iwls,
  title={IWLS 2005 benchmarks},
  author={Albrecht, Christoph},
  booktitle={International Workshop for Logic Synthesis (IWLS): http://www. iwls. org},
  year={2005}
}

@article{ghimire2025hardware,
  title={Hardware Design and Security Needs Attention: From Survey to Path Forward},
  author={Ghimire, Sujan and Chowdhury, Muhtasim Alam and Latibari, Banafsheh Saber and Mamun, Muntasir and Carpenter, Jaeden Wolf and Tan, Benjamin and Pearce, Hammond and Satam, Pratik and Salehi, Soheil},
  journal={arXiv preprint arXiv:2504.08854},
  year={2025}
}

@article{gubbi2024optimized,
  title={Optimized and automated secure ic design flow: A defense-in-depth approach},
  author={Gubbi, Kevin Immanuel and Latibari, Banafsheh Saber and Chowdhury, Muhtasim Alam and Jalilzadeh, Afrooz and Hamedani, Erfan Yazdandoost and Rafatirad, Setareh and Sasan, Avesta and Homayoun, Houman and Salehi, Soheil},
  journal={IEEE Transactions on Circuits and Systems I: Regular Papers},
  year={2024},
  publisher={IEEE}
}

@article{gubbi2023hardware,
  title={Hardware trojan detection using machine learning: A tutorial},
  author={Gubbi, Kevin Immanuel and Saber Latibari, Banafsheh and Srikanth, Anirudh and Sheaves, Tyler and Beheshti-Shirazi, Sayed Arash and PD, Sai Manoj and Rafatirad, Satareh and Sasan, Avesta and Homayoun, Houman and Salehi, Soheil},
  journal={ACM Transactions on Embedded Computing Systems},
  volume={22},
  number={3},
  pages={1--26},
  year={2023},
  publisher={ACM New York, NY}
}

@INPROCEEDINGS{logreg,
  author={Zou, Xiaonan and Hu, Yong and Tian, Zhewen and Shen, Kaiyuan},
  booktitle={2019 IEEE 7th International Conference on Computer Science and Network Technology (ICCSNT)}, 
  title={Logistic Regression Model Optimization and Case Analysis}, 
  year={2019},
  volume={},
  number={},
  pages={135-139},
  doi={10.1109/ICCSNT47585.2019.8962457}}

@article{ranfor,
title = {A comparison of random forest variable selection methods for classification prediction modeling},
journal = {Expert Systems with Applications},
volume = {134},
pages = {93-101},
year = {2019},
issn = {0957-4174},
doi = {https://doi.org/10.1016/j.eswa.2019.05.028},
url = {https://www.sciencedirect.com/science/article/pii/S0957417419303574},
author = {Jaime Lynn Speiser and Michael E. Miller and Janet Tooze and Edward Ip},
}

@article{dnn,
  author       = {Alfredo Canziani and
                  Adam Paszke and
                  Eugenio Culurciello},
  title        = {An Analysis of Deep Neural Network Models for Practical Applications},
  journal      = {CoRR},
  volume       = {abs/1605.07678},
  year         = {2016},
  url          = {http://arxiv.org/abs/1605.07678},
  eprinttype    = {arXiv},
  eprint       = {1605.07678},
  bibsource    = {dblp computer science bibliography, https://dblp.org}
}

@article{svm,
  title={Machine learning applications based on SVM classification a review},
  author={Abdullah, Dakhaz Mustafa and Abdulazeez, Adnan Mohsin},
  journal={Qubahan Academic Journal},
  volume={1},
  number={2},
  pages={81--90},
  year={2021}
}

\end{document}